\documentstyle[12pt]{article}
\newlength{\pubnumber} 

\catcode`\@=11
\@addtoreset{equation}{section}
\def\section{\@startsection{section}{1}{\z@}{3.5ex plus 1ex minus .2ex}
 {2.3ex plus .2ex}{\large\bf}}
\def\subsection{\@startsection{subsection}{2}{\z@}{2.3ex plus .2ex}
 {2.3ex plus .2ex}{\bf}}

\def\beq{\begin{equation}}
\def\eeq{\end{equation}}
\def\beqn{\begin{eqnarray}}
\def\eeqn{\end{eqnarray}}

\def\Str{{{\rm Str}\,}}

\def\inbar{\,\vrule height1.5ex width.4pt depth0pt}

\def\IC{\relax\hbox{$\inbar\kern-.3em{\rm C}$}}
\def\IQ{\relax\hbox{$\inbar\kern-.3em{\rm Q}$}}
\def\IR{\relax{\rm I\kern-.18em R}}
 \font\cmss=cmss10 \font\cmsss=cmss10 at 7pt
\def\IZ{\relax\ifmmode\mathchoice
 {\hbox{\cmss Z\kern-.4em Z}}{\hbox{\cmss Z\kern-.4em Z}}
 {\lower.9pt\hbox{\cmsss Z\kern-.4em Z}}
 {\lower1.2pt\hbox{\cmsss Z\kern-.4em Z}}\else{\cmss Z\kern-.4em Z}\fi}

\begin{document}

\begin{titlepage}
\samepage{
\setcounter{page}{1}
\rightline{IASSNS-HEP-95/42}
\rightline{McGill/95-28}
\rightline{\tt hep-th/9506001}
\rightline{June 1995}
\vfill
\begin{center}
 {\Large \bf Supertraces in String Theory\footnote{
         Talk given by R.C.M.\ at {\it Strings '95:  Future
          Perspectives in String Theory}, held
          at the University of Southern California,
          Los Angeles, CA, 13--18 March 1995.  To appear
          in the Proceedings published by World Scientific.}\\}
\vfill
 {\large Keith R. Dienes$^1$\footnote{E-mail address:
  dienes@guinness.ias.edu}, Moshe Moshe$^2$\footnote{E-mail address:
  phr74mm@vmsa.technion.ac.il}, and Robert C. Myers$^3$\footnote{
  E-mail address:  rcm@hep.physics.mcgill.ca}\\}
\vspace{.25in}
 {\it  $^1$  School of Natural Sciences, Institute for Advanced Study\\
  Olden Lane, Princeton, NJ  ~08540~  USA\\}
\vspace{.05in}
 {\it $^2$ Department of Physics, Technion -- Israel Inst.\ of Technology\\
  Haifa 32000, Israel\\}
\vspace{.05in}
 {\it $^3$ Department of Physics, McGill University \\
  3600 University St., Montr\'eal, Qu\'ebec  H3A-2T8  Canada\\}
\end{center}
\vfill
\begin{abstract}
  {\rm  We demonstrate that the spectrum of any consistent
   string theory in $D$ dimensions
   must satisfy a number of supertrace constraints:
   ${\rm Str}\,M^{2n}=0$ for $0\leq n<D/2-1$, $n\in \IZ$.
   Our results hold for a large class of string theories, including
   critical heterotic strings. For strings lacking spacetime supersymmetry,
   these supertrace constraints will be satisfied as a consequence of
   a hidden ``misaligned supersymmetry'' in the string spectrum.}
\end{abstract}

\vfill}
\end{titlepage}

\setcounter{footnote}{0}

Supertraces are of interest in quantum field theories because
they control the structure of the ultraviolet divergences appearing
in various loop amplitudes. For example, in four-dimensional spacetime,
the one-loop vacuum energy density (cosmological
constant) contains quartic, quadratic
and logarithmic divergences proportional to $\Str M^0$,
$\Str M^2$, and $\Str M^4$, respectively.
In this talk, supertraces are considered in the context of string theory,
and it will be shown that at tree level,
any consistent string theory satisfies
${\rm Str}\,M^{2n}=0$ for $0\leq n<D/2-1$,
$n\in \IZ$, where $D$ is the number of spacetime dimensions.
The details of this work may be found in Ref.~[1]. 

In calculating supertraces for string spectra, one faces an
immediate problem. In string theory, the physical spectrum contains
an {\it infinite}\/ number of states arranged
in towers whose levels are integer-spaced (in Planck-scale
units), and whose state degeneracies grow exponentially with mass.
Thus, to properly define the string-theoretic supertraces over the whole
string spectrum, we
regulate the sum over states as follows:
\beq
      \Str M^{2n}~\equiv~ \lim_{\gamma\to 0}\,
         \left\lbrace\sum_{\rm{physical}\atop{states}}\, (-1)^F\, (M_i)^{2n}
        \,e^{-4\pi\gamma M_i^2}\right\rbrace~.
\label{regulator}
\eeq
As is conventional, all masses are expressed in units of the Planck mass,
and hence $\gamma$ is a dimensionless quantity. (The factor
of $4\pi$ is introduced in the exponential as a convenient normalization
for what follows.)

In order to derive our supertrace constraints,
we exploit the remarkable expression for the string-theoretic
one-loop cosmological constant derived by Kutasov and
Seiberg \cite{kutsei}:
\beq
   \Lambda~=~ {\pi \over 3} \, \lim_{\gamma\to 0} \,
     (\gamma)^{1-D/2}\, \sum_{\rm{physical}\atop{states}} (-1)^F \,
        e^{-4\pi \gamma M_i^2 } ~.
\label{cosmik}
\eeq
In contrast to the orthodox one-loop formula, Eq.~(\ref{cosmik})
explicitly relates $\Lambda$ to the tree-level spectrum
of physical string states.
This formula applies for a large class of tachyon-free string theories
including all unitary non-critical strings, critical Type-II
strings, as well as the phenomenologically
interesting case of $D>2$ critical heterotic strings.

As stated above, Eq.~(\ref{cosmik}) is applicable for string
theories without physical tachyons. This is part of what defines
a physically consistent string theory, and the absence of tachyons
ensures that $\Lambda $ is free of infrared divergences.
Another remarkable feature of a consistent string theory is
modular invariance, a worldsheet symmetry of the one-loop
amplitudes.  For the present purposes, the most important
consequence of modular invariance is that the
one-loop amplitudes are ultraviolet finite. Thus string consistency
automatically ensures that $\Lambda $ is finite. Such a
result can only achieved in
Eq.~(\ref{cosmik}) if to leading order, as $\gamma\to 0$,
\beq
   \sum_{\rm{physical}\atop{states}} (-1)^F \,e^{-4\pi\gamma M_i^2 } ~\sim~
    \gamma^\delta ~~~{\rm with}~~ \delta \geq  D/2-1~.
\eeq

With this insight in hand,
it is straightforward to derive our supertrace
constraints by combining Eqs.~(\ref{regulator}) and (\ref{cosmik}):
\beqn
      \Str M^{2n}  &=&
        \lim_{\gamma\to 0}\,
         \left\lbrace\sum_i\, (-1)^F\, (M_i)^{2n}
        \,e^{-4\pi\gamma M_i^2}\right\rbrace \nonumber\\
     &=&
    \lim_{\gamma\to 0} \,\left\lbrace
    \left( {-1\over 4\pi}  {d\over d\gamma}\right)^n
   \sum_i \,(-1)^F \, e^{-4\pi \gamma M_i^2} \right\rbrace \nonumber\\
     &=&
    \lim_{\gamma\to 0} \,\left\lbrace
    \left( {-1\over 4\pi}  {d\over d\gamma}\right)^n
    \left\lbrack {3\over \pi}\,\Lambda ~ {\gamma}^{D/2-1} +\ldots
    \right\rbrack \right\rbrace
\eeqn
We see that the right side will vanish if $n<D/2-1$.
Thus we find that the spectra
of all consistent unitary non-critical strings and critical
Type-II and $D>2$ heterotic strings
must satisfy the supertrace constraints:\footnote{
     For $D=2$ critical heterotic strings, Eq.~(\ref{cosmik}) is
     modified to $\Lambda = {\pi \over 3} \lim_{\,\gamma\to 0}
     \,\lbrace\sum (-1)^F \, e^{-4\pi \gamma M_i^2 }\rbrace
     -8\pi\,N$, where $N$ is
     the number of bosonic minus fermionic unphysical tachyons with right- and
     left-moving squared masses (0,$-1$) respectively.
     This immediately yields the supertrace
     formula $\Str M^0={3\over\pi}\,\Lambda+24 \,N$, which may
     be compared to Eq.~(\ref{even}).}
\beq
      \Str\,M^{2n}~=~ 0~~~{\rm for}~~0\leq n< D/2-1, ~~n\in {\IZ}~.
\label{generalD}
\eeq
For an even number of dimensions, we also have
\beq
      \Str M^{D-2}  ~=~ {3\over \pi}\, {(D/2-1)!\over (-4\pi)^{D/2-1}}\,
      \Lambda  ~.
\label{even}
\eeq

For a string theory with spacetime supersymmetry (SUSY), these relations
are trivially satisfied through an exact boson/fermion degeneracy
at each mass level.  However, even in the absence of spacetime
SUSY, these constraints will still be satisfied.
These supertrace constraints are thus a generic
feature of tachyon-free tree-level string vacua, with or without
spacetime SUSY.

It is interesting to consider how the spacetime
bosons and fermions must be arranged throughout the mass levels
of non-SUSY string models
in order that Eqs.~(\ref{generalD}) and (\ref{even}) are satisfied.
As an explicit example of a tachyon-free, modular-invariant,
non-SUSY string theory, therefore, let us consider the
$SO(16)\times SO(16)$
string in ten dimensions \cite{osixteen}. In this case, we know that
$\Str\,M^0$, $\Str\,M^2$, $\Str\,M^4$, and $\Str\,M^6$ must all vanish,
and $\Str\,M^8\approx-0.668$ (where
we have used the numerical value of $\Lambda$ for this model \cite{osixteen}).
At the massless level, one finds a surplus of
2112 fermionic states over bosonic states.
At the first massive level, though, there is a surplus of 147,456 bosonic
states,
which more than compensates for the previous surplus of fermions.
This surplus of bosons is then (over-)compensated by a surplus of
4,713,984 fermions at the next level.
In fact, as shown in Fig.~1, this pattern of
carefully balanced boson/fermion oscillations
continues throughout the entire infinite tower of massive string states.

These boson/fermion oscillations turn out to be
the signature of a hidden so-called ``misaligned supersymmetry''
which generically appears in the spectra of all tachyon-free
modular-invariant string theories in all dimensions,
and which arises even in the absence of full spacetime SUSY.
Misaligned SUSY is discussed in Ref.~[4].
Thus, we see that misaligned SUSY is the mechanism which allows
our supertrace constraints to be satisfied without spacetime
SUSY.  Indeed, rather than having cancellations within each multiplet
(as in ordinary softly or spontaneously broken SUSY field
theories), in string theory these supertrace constraints are satisfied
via cancellations between states at different energy levels across
the entire infinite string spectrum.
This suggests a possible new string-inspired mechanism for achieving
finiteness in field theory, and for constructing a possible non-SUSY
solution to the gauge hierarchy problem.

\vfill\eject

\input epsf
\begin{figure}[t]
\centerline{\epsfxsize 6 truein \epsfbox {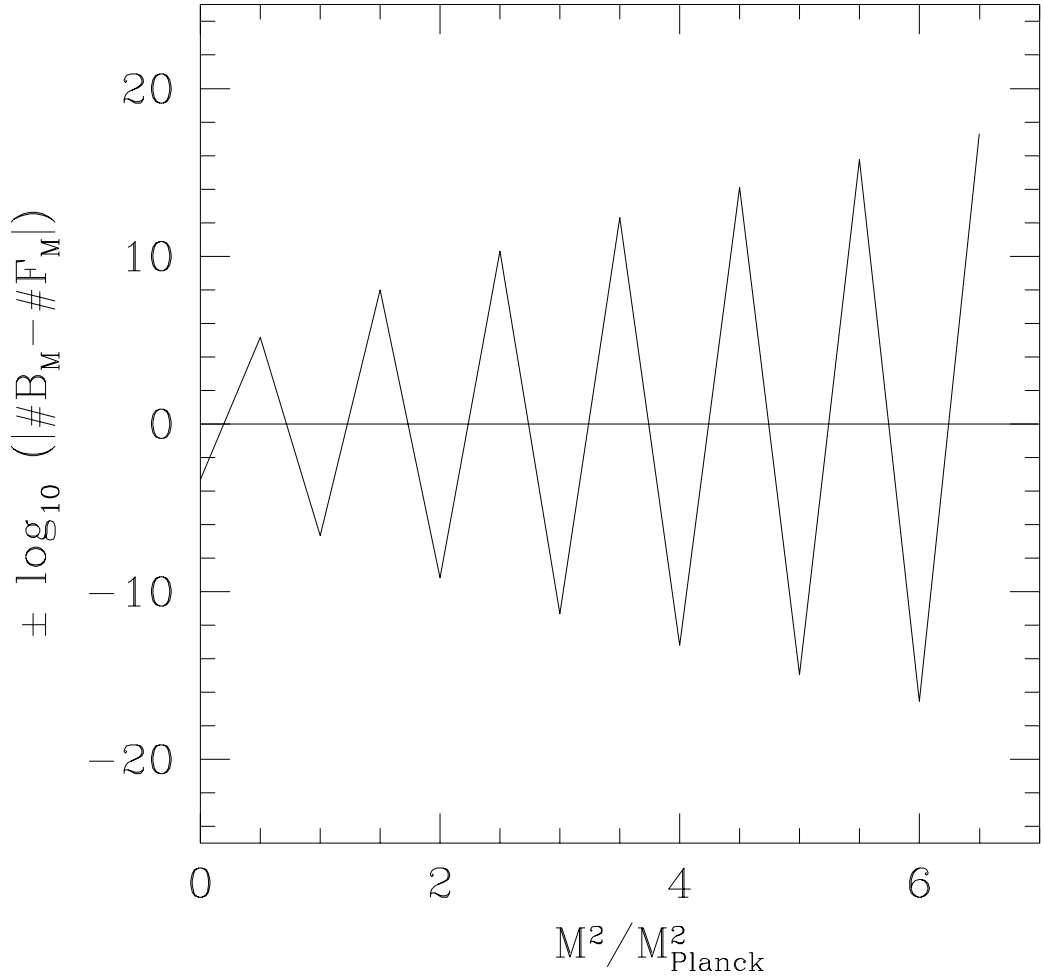}}
\nobreak
\caption{Boson/fermion oscillations in the $D=10$ non-supersymmetric
   tachyon-free $SO(16)\times SO(16)$ heterotic string.
   For each value of spacetime mass-squared (either integer or half-integer
   in this model), we have calculated the corresponding number of spacetime
   bosons minus fermions, and have plotted the logarithm
   of the absolute value of this difference (with an overall
   sign reflecting whether the difference is positive or negative ---
   {\it i.e.}, whether there are more bosons or fermions at that level).
   We have connected these points in order to stress the oscillatory
   behavior of the boson and fermion surpluses.  These oscillations
   insure that $ \Str\,M^0= \Str\,M^2= \Str\,M^4= \Str\,M^6=0$
   in this model, even though there is no spacetime supersymmetry.  }
\vskip -5 truein       
\end{figure}

 ~
\vfill\eject

\leftline{\large \bf Acknowledgments}
  \medskip
  \smallskip
R.C.M.\ thanks the organizers of Strings '95 for the opportunity
to present this talk.
This work was supported in part by DOE Grant No.\ DE-FG-0290ER40542,
the US/Israel Bi-National Science Foundation, the Technion VPR Fund,
NSERC (Canada), and FCAR (Qu\'ebec).

\bigskip
\medskip
\bibliographystyle{unsrt}

\begin{thebibliography}{99}
\bibitem[1]{supertrace}
     K.R. Dienes, M. Moshe, and R.C. Myers, {\tt hep-th/9503055},
     {\it Phys.\ Rev.\ Lett.}\/ {\bf 74} (1995) 4767.
\bibitem[2]{kutsei}
     D. Kutasov and N. Seiberg, {\it Nucl.\ Phys.}\/ {\bf B358} (1991) 600.
\bibitem[3]{osixteen}
     L. Dixon and J. Harvey, {\it Nucl.\ Phys.}\/ {\bf B274} (1986) 93;\\
     L. Alvarez-Gaum\'e, P. Ginsparg, G. Moore, and C. Vafa, {\it Phys.
     Lett.}\/ {\bf B171} (1986) 155.
\bibitem[4]{misaligned}
     K.R. Dienes, {\tt hep-th/9402006},
          {\it Nucl.\ Phys.}\/ {\bf B429} (1994) 533.\\
     A non-technical introduction can also be found in:\\
     K.R. Dienes, {\tt hep-th/9409114}  (to appear in the Proceedings of
              PASCOS '94);\\
     K.R. Dienes, {\tt hep-th/9505194} (to appear in the Proceedings of
              Strings '95).
\end{thebibliography}

\vfill\eject

\end{document}